\documentclass[conference,compsoc]{IEEEtran}

\usepackage[dvipsnames, table]{xcolor}

\usepackage{url}
\usepackage[utf8]{inputenc}
\usepackage{graphicx}
\usepackage{hyperref}
\usepackage{amsfonts} 
\usepackage{amsmath}
\usepackage{booktabs}
\usepackage{fancyhdr}
\usepackage{listings}
\usepackage{float}
\usepackage{xspace}
\usepackage{adjustbox}
\usepackage{mathtools}
\usepackage{textcomp}
\usepackage{boldline}
\usepackage{import}
\usepackage{capt-of}
\usepackage{pifont}

\usepackage{nicematrix}
\usepackage{multirow}
\usepackage{mdframed}

\usepackage[frozencache=true,cachedir=minted]{minted}

\usepackage[capitalize]{cleveref}

\usepackage[backend=biber,isbn=false,url=false,doi=false,style=numeric,sorting=none,maxbibnames=99,giveninits=true]{biblatex}

\newcommand{\algowner}{CAO\xspace}
\newcommand{\cloud}{SP\xspace}

\newcommand{\instrmeas}{IM\xspace}
\newcommand{\instrmeass}{IMs\xspace}
\newcommand{\candidate}{candidate\xspace}
\newcommand{\Candidate}{Candidate\xspace}

\newcommand{\baseline}{native system\xspace}

\newcommand{\BaselinE}{Native System\xspace}
\newcommand{\interpreted}{WASM system\xspace}
\newcommand{\interpreteD}{WASM System\xspace}

\newcommand{\nativecomp}{native~execution\xspace}
\newcommand{\Nativecomp}{Native~execution\xspace}
\newcommand{\ircomp}{IR~execution\xspace}

\newif\ifsubmission{}
\submissionfalse

\newif\ifprintheader
\printheaderfalse

\newif\ifdraft{}

\drafttrue{}

\draftfalse{}

\ifdraft

\printheadertrue

\newcommand{\todo}[1]{\textcolor{red}{TODO: #1}}

\newcommand{\ivan}[1]{\textbf{\emph{ #1 \colorbox{magenta}{[Ivan]}}}}
\newcommand{\moritz}[1]{\textbf{\emph{ #1 \colorbox{yellow}{[Moritz]}}}}
\newcommand{\srdjan}[1]{\textbf{\emph{ #1 \colorbox{blue}{\textcolor{white}{[Srdjan]}}}}}

\else

\newcommand{\todo}[1]{}

\newcommand{\ivan}[1]{}
\newcommand{\moritz}[1]{}
\newcommand{\srdjan}[1]{}

\fi

\ifsubmission
\printheaderfalse
\fi

\renewcommand{\paragraph}[1]{\vspace{0.5em}\noindent\textbf{#1.}}
\newcommand{\bfparagraph}[1]{\paragraph{#1}}

\newcommand{\Yes}[0]{\ding{51}}
\newcommand{\No}[0]{\ding{55}}

\let\oldding\ding%
\renewcommand{\ding}[2][1]{\scalebox{#1}{\oldding{#2}}}

\newcommand{\five}{\ding[1.2]{196}\xspace}

\definecolor{lispgreen}{RGB}{154, 228, 151}

\lstdefinelanguage{myasm}{
    language=[x86masm]Assembler,
    basicstyle=\ttfamily\small,
    morekeywords={movswq, jmpq},
    sensitive=false,
    morecomment=[l]{//},
    morecomment=[l]{\#},
    morecomment=[s]{/*}{*/},
    commentstyle=\color{gray},
    morestring=[b]",
    classoffset=1,
    alsoletter={\%},
    keywords={\%rax,\%rbx,\%rdx,\%rcx,\%cl,\%eax,\%ebx,\%ecx},
    keywordstyle=\color{orange},
    classoffset=2,
    keywords={\%rbp},
    keywordstyle=\color{purple},
    classoffset=0,
    numbers=left,
    numberstyle=\ttfamily\small,
}

\lstdefinelanguage{myC}{
    language=[ANSI]C,
    keywordstyle=\color{magenta},
    basicstyle=\ttfamily\small,
    breaklines=true,
    sensitive=false,
    morecomment=[l]{//},
    morecomment=[l]{\#},
    morecomment=[s]{/*}{*/},
    commentstyle=\color{gray},
    morestring=[b]",
    otherkeywords={<,>,=,!=,*,++},
    emph={skip_label, emit_label,POP_I32,RESET_STACK,PRESERVE_LOCAL_FOR_BLOCK,wasm_loader_prepare_bytecode}, emphstyle=\color{LimeGreen},
    emph={[2]WASM_OP_UNREACHABLE,WASM_OP_NOP,WASM_OP_IF},                                                    emphstyle={[2]\color{RoyalPurple}},
    emph={[3]bool},                                                                                          emphstyle={[3]\color{Aquamarine}\itshape},
    stringstyle=ttfamilycolor{red!50!brown},
}

\newminted[myc]{pygmentize_customizations/custom_lexers.py:MyCLexer -x}{linenos,style=customc,numbersep=2pt}
\newminted[myasm]{pygmentize_customizations/custom_lexers.py:MyASMLexer -x}{linenos,style=customasm,numbersep=2pt}

\newmintedfile[cmintedfile]{pygmentize_customizations/custom_lexers.py:MyCLexer -x}{linenos,style=customc,numbersep=2pt}
\newmintedfile[asmmintedfile]{pygmentize_customizations/custom_lexers.py:MyASMLexer -x}{linenos,style=customasm,numbersep=2pt}

\immediate\write18{bash pygmentize_customizations/patch_pyg.sh}
\renewcommand{\MintedPygmentize}{./pygmentize}

\makeatletter
\let\@float@c@listing\@caption
\makeatother

\DeclareSourcemap{
    \maps[datatype=bibtex]{
        \map[overwrite]{
            \step[fieldsource=doi, final]
            \step[fieldset=url, null]
            \step[fieldset=eprint, null]
        }
        \map[overwrite]{
            \step[fieldsource=note, final]
            \step[fieldset=year, null]
        }
        \map[overwrite]{
            \step[fieldsource=url, final]
            \step[fieldset=year, null]
        }
    }
}

\DeclareSourcemap{
    \maps[datatype=bibtex]{
        \map{
            \pertype{inproceedings}
                \step[fieldsource=series, fieldtarget=booktitleaddon]
        }
        \map{
            \pertype{proceedings}
                \step[fieldsource=series, fieldtarget=booktitleaddon]
        }
    }
}

\renewbibmacro*{booktitle}{%
    \ifboolexpr{
        test {\iffieldundef{booktitle}}
        and
        test {\iffieldundef{booksubtitle}}
    }
    {}
    {
        \iffieldundef{booktitleaddon}{
            \printtext[booktitle]{%
            \printfield[titlecase]{booktitle}%
            \setunit{\subtitlepunct}%
            \printfield[titlecase]{booksubtitle}}%
            \newunit%
        }%
        {
            \printtext[booktitle]{%
            \printfield[titlecase]{booktitle}%
            \addspace%
            \printtext[parens]{%
            \printfield[titlecase]{booktitleaddon}}%
            \setunit{\subtitlepunct}%
            \printfield[titlecase]{booksubtitle}}%
            \newunit%
        }%
    }
}

\renewbibmacro{in:}{}

\AtEveryBibitem{%
    \clearlist{address}
    \clearfield{date}
    \clearfield{pages}
    \clearlist{location}
    \clearfield{month}
    \clearfield{series}
    \clearfield{volume}
    \clearfield{number}
 
    \ifentrytype{book}{}{%
        \clearlist{publisher}
        \clearname{editor}
    }
    \ifentrytype{InProceedings}{
        \clearfield{year}
    }{}
    \ifentrytype{inproceedings}{
        \clearfield{year}
    }{}
    \ifentrytype{INPROCEEDINGS}{
        \clearfield{year}
    }{}
    \ifentrytype{incollection}{
        \clearfield{year}
    }{}
}

\title{On (the Lack of) Code Confidentiality in Trusted Execution Environments}

\author{ }
\author{Ivan Puddu, Moritz Schneider, Daniele Lain, Stefano Boschetto, Srdjan \v{C}apkun\\
    Department of Computer Science\\
    ETH Zurich\\
    \{name.surname\}@inf.ethz.ch}

\begin{document}

\ifprintheader
    \renewcommand{\headrulewidth}{0pt}
    \fancyhf{}%
    \chead{\textcolor{red}{Confidential Draft -- Do NOT distribute!!}}
\else
    \ifsubmission
        \author{}
    \fi
\fi

\ifprintheader
    \thispagestyle{fancy}
\fi

\maketitle

\begin{abstract}
Trusted Execution Environments (TEEs) have been proposed as a solution to protect code confidentiality in scenarios where computation is outsourced to an untrusted operator. We study the resilience of such solutions to side-channel attacks in two commonly deployed scenarios: when a confidential code is a native binary that is shipped and executed within a TEE and when the confidential code is an intermediate representation (IR) executed on top of a runtime within a TEE.
We show that executing IR code such as WASM bytecode on a runtime executing in a TEE leaks most IR instructions with high accuracy and therefore reveals the confidential code. Contrary to IR execution, native execution is much less susceptible to leakage and largely resists even the most powerful side-channel attacks.
We evaluate native execution leakage in Intel SGX and AMD SEV and experimentally demonstrate end-to-end instruction extraction on Intel SGX, with WASM bytecode as IR executed within WAMR, a hybrid between a JIT compiler and interpreter developed by Intel. Our experiments show that IR code leakage from such systems is practical and therefore question the security claims of several commercial solutions which rely on TEEs+WASM for code confidentiality.

\end{abstract}

\section{Introduction}

The trend of outsourcing data storage and computation has given rise to concerns about the confidentiality of not only data but also of code that is running on remote (typically cloud) services. To address these broad concerns, confidential computing, based on Trusted Execution Environments (TEEs) such as Intel SGX~\cite{intelsgx} and AMD SEV~\cite{amdsev}, has been deployed in today's commercial cloud~\cite{googlecloud_sev, azure_sgx, azure_sev}. %

TEEs allow the client to deliver their confidential code and data into a protected CPU enclave, which then isolates it from the OS and hypervisor that are running on the same machine and, more generally, from the untrusted Service Provider (SP). This is typically achieved via attestation - the client first sends the public part of its code to the SP (e.g., a VM), attests that this code is running within an enclave, establishes a secure channel (typically TLS) to the enclave, and then uses the secure channel to deliver confidential code and data into the enclave. Once the confidential code is delivered to the enclave, it can be executed in isolation. Recent years have seen the emergence of several designs that generally follow this approach, use different TEEs, and offer various trade-offs, both as academic proposals~\cite{vc3, sgxelide, teeshift, obfuscuro, twine, trustjs, acctee, heaven} and commercial solutions~\cite{enarx, veracruz, edgeless, scone}. 

\begin{figure}[t]
    \centering
    \includegraphics[trim={0 0 0 0},clip,width=\linewidth]{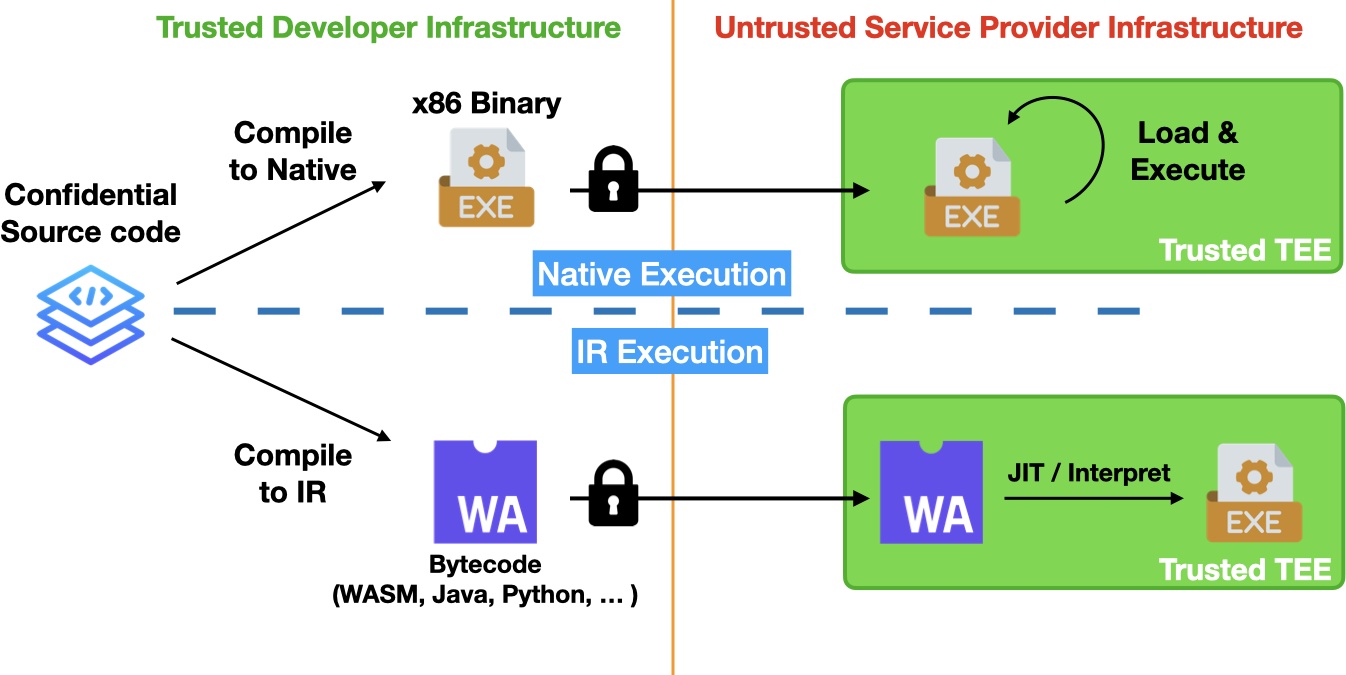}
    \vspace{-15pt}
    \caption{The two main approaches used to provide code confidentiality with TEEs: \nativecomp (above the dashed line) and \ircomp (below the dashed line). }
    \label{fig:both_compositions_overview}
\end{figure}

One of the core ways in which these solutions diverge is the format in which the confidential code is delivered to the enclave. They typically follow one of two approaches: \emph{\nativecomp} and \emph{\ircomp}, where IR stands for Intermediate Representation. We illustrate these approaches in \cref{fig:both_compositions_overview}. %
In \nativecomp, the developer compiles the confidential code to a \emph{native} binary (i.e., x86) and then, after initializing a remote enclave, sends the binary to it. In \ircomp, the developer compiles the confidential code to bytecode (e.g., WASM or Java) or directly sends the source code to the enclave. Whereas in the case of native execution, the enclave can simply copy the instructions from the received binary to memory and execute them, in the case of IR execution, it needs first to convert the received IR into native code. This is done by a Virtual Machine (VM)-like environment in which either a just-in-time (JIT) compiler first converts the IR code to native or an interpreter directly executes it. A number of academic and commercial systems now support either native or IR execution within TEEs. WASM runtimes are particularly well supported~\cite{enarx,veracruz,edgeless,twine,acctee} because WASM requires a small runtime resulting in a small TCB. Moreover, more than 40 programming languages can currently be compiled to WASM, with support for more underway~\cite{webassembly_support}.

However, even if several~~\cite{vc3, sgxelide, teeshift, obfuscuro, trustjs, heaven, enarx, veracruz, scone} of these systems claim to support code confidentiality for native or \ircomp, so far, these claims have not been evaluated in the open literature.

\paragraph{Our paper} We perform the first analysis of confidential code leakage in native and \ircomp of modern TEEs. In particular, we evaluate code leakage on \nativecomp from Intel SGX and AMD SEV TEEs (x86 ISA) and \ircomp with WASM runtimes. %
In our evaluation, we single-step the enclaves by controlling interrupts and record various side-channel measurements for each instruction. This allows building a trace of the execution of the victim enclave at the instruction granularity in an attempt to identify individual IR instructions or instruction sequences.

Our results show that \nativecomp is largely robust to even the most sophisticated side-channel attacks and leaks limited information about individual instructions. 
\ircomp, which we tested on WAMR~\cite{wamr}, a lightweight WASM interpreter developed by the Bytecode Alliance, however, has shown to be highly vulnerable to our side-channel analysis. We successfully leaked more than 45\% of the secret instructions with 100\% confidence from a synthetic C program running various math and cryptographic functions and from a chess engine written in Rust~\cite{rust_chess}. Collectively, we successfully extracted over 1 billion WASM instructions from both code sets, albeit not all with 100\% confidence. This is possible because we are able to leak around 80\% of the instructions in the WASM instruction set architecture (ISA) with 100\% confidence. This level of confidence is obtained from just one run of the victim enclave. 

These results are consistent with the expected side-channel leakage. Each IR instruction is represented by a number of native instructions. To identify an IR instruction, the attacker can therefore rely on a much longer side channel trace than when it tries to identify an individual native instruction. Therefore, it is clear that \ircomp will always be more vulnerable to code leakage than \nativecomp. Our results show that in the case of \ircomp, such leakage is also practical, which raises questions about the security guarantees of any \ircomp in TEEs.

\bfparagraph{Contributions}
We summarize our contributions as follows:
\begin{itemize}
    \item To our knowledge, this is the first study to investigate and bring forth the challenges in providing \emph{code} confidentiality in TEEs.
    \item We generalize system designs aiming to provide code confidentiality in TEEs into two, \nativecomp and \ircomp, and develop a methodology to quantify and compare their code leakage. %
    \item We analyze instruction leakage in both systems on various microarchitectures supporting TEEs from Intel and AMD. We show that, based on side-channels, extracting unknown code in \nativecomp can be considered out of reach even for an ideal attacker. On the other hand, \ircomp greatly amplifies any leakage from \nativecomp and allows us to extract most of the confidential instructions \emph{from a single execution}. 
    \item To demonstrate the practicality of these findings in \ircomp, we develop a practical end-to-end instruction extraction attack against WAMR, a WASM runtime running on Intel SGX.
\end{itemize}

\paragraph{Responsible disclosure}
On 02 November 2022, we disclosed our findings to the following companies promising code confidentiality in TEEs: Veracruz (ARM)~\cite{veracruz},  Edgeless~\cite{edgeless}, Enarx~\cite{enarx}, and Scone~\cite{scone}. Edgeless acknowledged receiving our paper but did not take any further steps. Enarx responded that they are researching mitigations, while Scone told us that they are working on mitigating the reported issues. Veracruz responded that side-channels are out of scope in their attacker model. Nonetheless, they are working on clarifying their documentation about the risks related to code confidentiality in TEEs.

\section{System and Attacker Model}\label{sec:model}

We consider a setting in which computation is outsourced while needing to safeguard the confidentiality of the code used for computation. Two main parties are involved in this setting: 
\begin{itemize}
    \item A \textit{Confidential Algorithm Owner} (\algowner) that wants to offload computation to the cloud while keeping their code confidential; and
    \item A \textit{Service Provider} (\cloud) that provides support for Trusted Execution Environments (TEEs).
\end{itemize}

While the \cloud TEEs provide memory confidentiality at runtime, the \algowner cannot simply create an \emph{enclave} (a TEE instance) containing the confidential code, ship it to the \cloud and expect it to remain confidential: on both Intel SGX and AMD SEV, the initial state of the enclave is visible by the untrusted operating system (OS) and/or the hypervisor. Academic~\cite{vc3, sgxelide, teeshift, obfuscuro, twine, trustjs, acctee, heaven} and industrial~\cite{enarx, veracruz, edgeless, scone} solutions address this problem by supplying the confidential code to the enclave only after the enclave has been initialized and attested. The confidential part of the code is, therefore, only communicated to the enclave after the attestation and the creation of the secure channel between the \algowner and the enclave. 

Typically, two main approaches are employed to supply and execute confidential code in an enclave: \emph{\nativecomp} and \emph{\ircomp}. Each can be further broken down into three stages: (i) compile, (ii) attest, and (iii) deploy and execute.

\paragraph{(i) Compile}
In this stage, the \algowner compiles its confidential source code for the TEE. 
\Nativecomp approaches~\cite{vc3, sgxelide, teeshift, obfuscuro, scone, edgeless, heaven} require the \algowner to compile to a native format (we focus on x86 object binaries). In \ircomp~\cite{enarx, veracruz, scone, edgeless, twine, trustjs, acctee}, code gets compiled to an \textit{intermediate representation} (IR) chosen as a compilation target, e.g., WebAssembly (WASM) bytecode, Javascript, Python, or Go. In some of the systems, the compilation step is skipped as the TEE directly interprets the source code.

\paragraph{(ii) Attest}
In this stage, the \algowner deploys an initial, non-confidential code with the SP and attests that this code is initialized in the enclave. Attestation ensures that the initial enclave has been deployed in a legitimate TEE and that its integrity is guaranteed. This initial enclave code is often provided by the chosen framework or \cloud~\cite{scone, enarx, edgeless}.
As part of attestation, the \algowner bootstraps a secure channel (e.g., TLS) with the enclave. On this secure channel, the \algowner sends either the confidential code to the enclave or a key to decrypt a confidential code image already contained in the initial enclave.

\paragraph{(iii) Deploy and Execute}
After the attestation, the \algowner instructs the initial enclave to execute the confidential code. In \nativecomp, this is straightforward - the enclave simply jumps to the entry point of the x86 confidential code, which was stored in its memory as a result of the previous stage.
In \ircomp, the initial enclave contains an interpreter (e.g., WASM or Python), potentially with a just-in-time (JIT) compiler; the confidential instructions get interpreted, and if a JIT compiler is available, some parts get compiled to native (x86) to speed up the execution.

\subsection{Attacker Model}\label{sec:att_model}
The goal of the attacker is to leak the instructions and, therefore, the confidential code that is executing in the TEE. 
Here we assume that the attacker is either the Service Provider (\cloud) or has privileged access to the server in which the confidential code is executing, i.e., the attacker controls the supervisor software, that is, the hypervisor (on a system with AMD SEV) and/or the operating system (for Intel SGX). This is a standard attacker model for TEEs~\cite{sgxexplained,amdsevsnp}. The attacker can see the non-confidential, initial enclave code as this code is provided in cleartext to the OS and hypervisor to load the enclave; typically, this code is public. We assume that the attacker has no control over when the confidential algorithm is executed and which secret inputs are given to it. This assumption impacts the side-channels available to the attacker, as, for instance, in this setting it is unrealistic to i) restart an enclave a large number of times to average out noise and ii) correlate the instructions across multiple runs - as different code paths might be executed depending on the supplied inputs.

Since the attacker has control over supervisor software on the system, they are able to: manipulate interrupts, observe changes to paging management structures (such as page table entries), and other information available to the OS, such as the last branch record (LBR). These capabilities\footnote{As demonstrated in the literature against SGX~\cite{sgx-step, nemesis, accesscontrolledchannel,leakycaulderon17,lee2017inferring,moghimi2020copycat}; they apply to AMD SEV as well, as discussed in \cref{sec:discussion}.} allow the attacker to single-step the TEE execution (through interrupts), see whether memory read and writes are executed (through the page tables), the approximate location (down to the cacheline) of memory read and writes, which code-page is being executed, whether some types of jumps were executed, and the execution time of interrupted instructions. %
We refer to an attacker with these capabilities as the \emph{state-of-the-art} (SotA) attacker. %

Throughout the paper, unless otherwise specified, we employ a SotA attacker. However, when necessary to establish upper bounds on code leakage, we use a stronger attacker model, which we refer to as the \emph{ideal attacker}. As the ideal attacker is specific to the system for which we want to estimate an upper bound, we will only introduce it when needed in the following sections.

\section{Leakage Analysis Overview}\label{sec:overview}

\begin{figure}[t!]
    \centering
    \includegraphics[trim={175 150 175 95},clip,width=\linewidth]{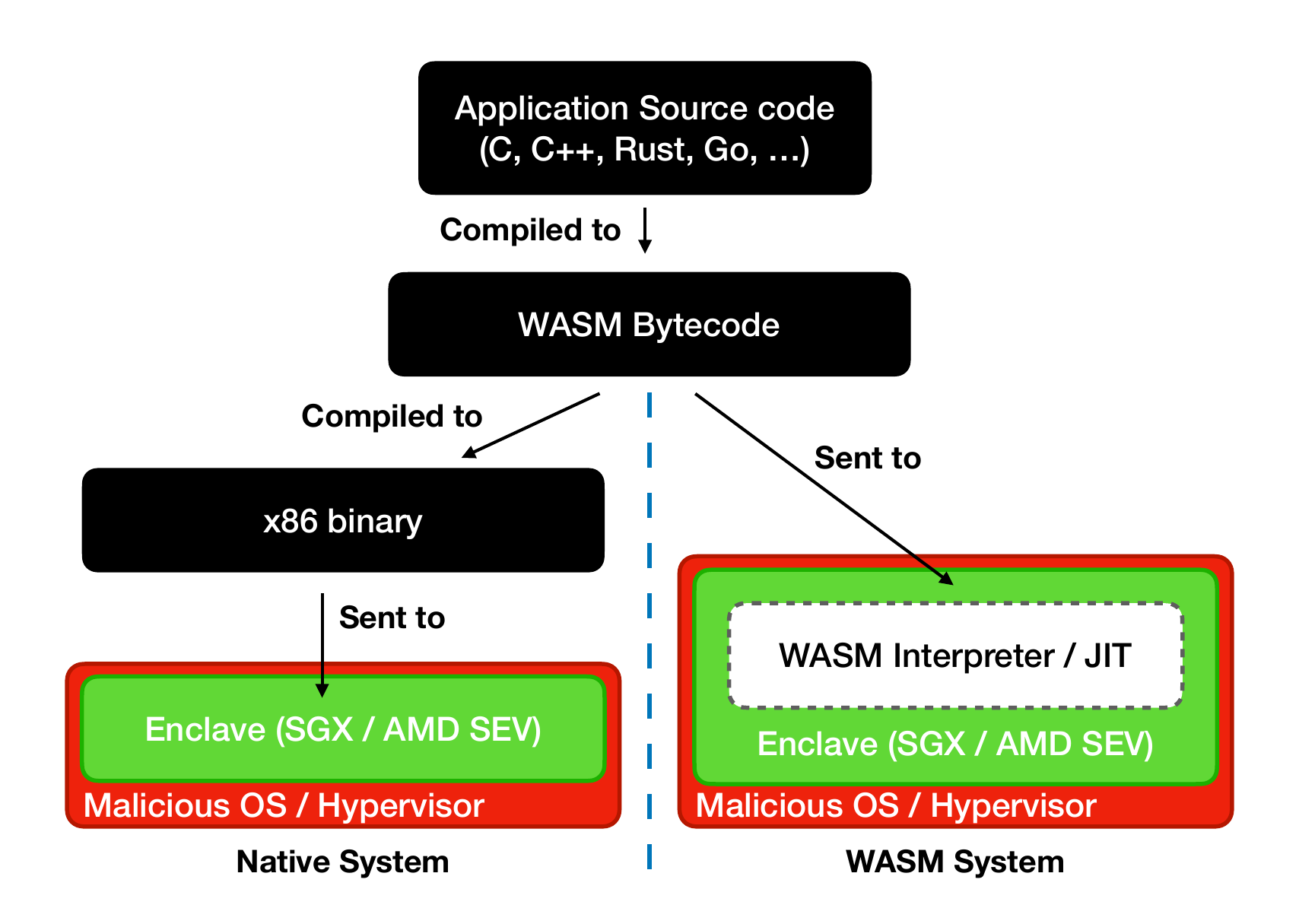}
    \vspace{-20pt}
    \caption{The two approaches to code confidentiality in TEEs. The \nativecomp enclave (left) gets the source code compiled to x86; the \ircomp enclave (right) gets as input WASM bytecode. Both systems operate in an environment with a malicious OS and are tasked with executing the same source code.}
    \label{fig:att_overview}
    \vspace{-1em}
\end{figure}

To compare the leakage in native and \ircomp, we instantiate them in two systems, the \emph{\baseline} and the \emph{\interpreted} illustrated in \cref{fig:att_overview}. %
The \emph{\baseline} accepts and executes confidential instructions in x86 (native) binary format. %
The \emph{\interpreted} implements \ircomp by accepting as input WASM bytecode instructions. The \interpreted enclave can then either interpret the bytecode or process it with a JIT compiler before execution. We refer, in general, to interpreters and JIT compilers as \emph{translators}.
We choose WebAssembly (WASM) to evaluate intermediate representation (IR) leakage due to its widespread adoption, large language support (more than 40 languages can be compiled to WASM bytecode~\cite{webassembly_support}), and the existence of multiple stable and lightweight runtimes.
Further, it can easily be compiled into native code, making the comparison between the two systems easier and more rigorous.
The enclaves in the two systems get the instructions in different formats from the same source program. We compile the source code to WASM bytecode and then the bytecode to x86 outside the enclave (cf. \cref{fig:att_overview}). The \baseline is given the final x86 binary, while the \interpreted is given the intermediate WASM bytecode.
Thus, the two systems are tasked with executing the very same program, allowing us to attribute any possible differences in leakage to the system running the instructions.

There are two fundamental differences between the \baseline, and \interpreted that influence their susceptibility to side-channels: (i) translators often execute more low-level instructions than equivalent native binaries, and (ii) the instruction set architectures (ISAs) of native instructions are usually considerably bigger than the ISAs used for interpreted languages.
Combining these two observations, our hypothesis is that the \interpreted is potentially more leaky than the \baseline due to having longer (and thus more unique) patterns of execution traces and having fewer possible instructions in the ISA that generate these traces.
In the following, we expand on these differences.

\begin{figure}[tb]\centering\sffamily\scriptsize
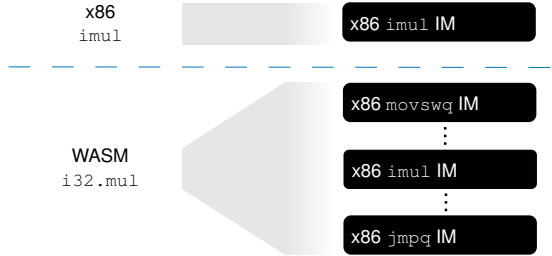
\caption{Sample trace collection during the execution of a x86 \texttt{imul} and one of its WASM equivalents, \texttt{i32.mul}. The x86 instruction generates a single Instruction Measurement (IM), while the WASM instruction generates 9 IMs due to executing 9 underlying x86 instructions.}\label{fig:amplification}
\end{figure}

\paragraph{Number of executed native instructions}
Translators of high-level languages with powerful semantics execute multiple native instructions for each high-level instruction.
These translators thus \emph{amplify} the amount of information an attacker can collect during the execution of interpreted code, compared to attacking a \baseline. 
For example, \cref{fig:amplification} shows the difference in collected traces by an attacker when profiling one x86 instruction versus one of its equivalents in a WASM interpreter.

All translators, from high-performance JIT-based to interpreters, must perform two steps to execute a binary: first, they have to parse the code and, second, execute it. Parsing usually involves looping over each instruction of the input code, decoding it, and preparing it for execution (e.g., with a \texttt{switch-case} statement as shown in \cref{lst:wasm_loader}). As the underlying architecture does not provide single complex instructions to perform these operations, multiple native instructions are executed while parsing a single WASM instruction. Not only this, but since different WASM instructions require different actions by the parser, the amplified instructions will differ based on which WASM instruction is being parsed. Effectively, this creates an exploitable control-flow dependency.
Similar issues arise during execution. For instance, the WASM \texttt{add} instruction adds the last two values from the WASM stack and then writes the result back to the stack. An interpreter needs first to read these values and then write the result back, generally using multiple native instructions for this task. In contrast, on x86, it is possible to perform all these operations with a single \texttt{add}.
In summary, the WASM system enclave executes several, and different, x86 instructions for each WASM instruction \emph{both during parsing and execution}.

While \cref{lst:wasm_loader} shows the implementation of the loader for the WAMR interpreter~\cite{wamr}, other WASM projects we inspected (Wasmtime~\cite{wasmtime} and Wasmer~\cite{wasmer}) have similar implementations. In fact, we remark that the amplification described above with the related control-flow dependency on input instructions is likely to be found in any interpreter or compiler available today. %
However, different implementations will exhibit different amplification factors, as, compared to each other, they might employ a different number of x86 instructions to parse and emulate high-level instructions. This aspect is crucial as it affects the exploitability of the high-level instructions.

\paragraph{Difference in ISAs}
The WASM Instruction Set Architecture (ISA) is significantly smaller than the x86 ISA (between $\approx$6x and $\approx$14x, depending on the x86 microarchitecture). Since the attacker knows that the enclave accepts only valid instructions, the attacker has fewer instructions to guess from in the \interpreted than in the \baseline. 
To give a concrete example of why this helps the attacker, consider the \texttt{add} instruction in x86 and WASM. In the WASM case, it can only add the two most recent values in the stack, while in the x86, many variations are possible, e.g., adding from different locations in memory, from registers, or even vectors. 
Assuming an attacker that can only leak the opcode (i.e., an \texttt{add}), this reveals more information in the \interpreted than in the \baseline.

\begin{listing}
    \cmintedfile{figures/code_examples/switch_case.c}
    \caption{Excerpt of the main loop of the Bytecode alliance WAMR interpreter~\cite{wamr} (commit \texttt{b554a9d}) responsible for loading a WASM binary. \texttt{opcode} (line $5$) is the opcode of the current WASM instruction being parsed. This listing shows how a control-flow dependency on the \texttt{opcode} usually manifests (line $8$) in WASM interpreters and compilers, and how different instructions exhibit different amplification factors. For instance, \texttt{WASM\_OP\_IF} (line $14$) requires multiple operations to be translated, amplifying the information available to the attacker compared to the equivalent functionality in x86 (usually a single instruction).}%
    \label{lst:wasm_loader}
\end{listing}

\section{Methodology}~\label{sec:methodology}
In our study, we single-step the enclave to collect information about each executed native instruction. We refer to the information collected for each native instruction as \emph{instruction measurement} (\instrmeas). Given the side-channels available in our attacker model, each \instrmeas contains the following information about an executed instruction: the execution time, the set of accessed code pages, the set of accessed data pages, and for each of the data pages, whether the access was a memory read or write. A series of \instrmeass forms an \emph{execution trace} containing all the information available to the attacker. Note that the trace contains as many \instrmeass as the x86 instructions measured. Thus, in the \baseline, there is one \instrmeas per confidential x86 instruction that the attacker wants to leak. On the other hand, in the WASM system, multiple \instrmeass are collected for each confidential WASM instruction. Finally, we can only measure instructions if they are executed; hence the execution trace only contains \instrmeass related to the executed branches and no information about non-executed code paths. 

\paragraph{Features}
It's worth noting that not all the information in an \instrmeas can be directly used to infer which instruction was executed. This is due to two reasons: first, the measurement might be too noisy, and second, it might be only related to an instruction's inputs and not to its operand. For instance, the side-channel used to measure the execution time is subject to noise, and it is, therefore, generally hard to discriminate instructions based on this measurement: a memory read (\texttt{mov}) and an addition from memory (\texttt{add}) are two very different instructions (in terms of a program's logic) that produce similar timing distributions~\cite{abel19a}. Thus, based on the timing information alone, an attacker would not be able to distinguish between the two. With respect to the second reason, knowing the data page that was accessed does not generally contain any information about the instruction type - the relevant piece of information about the instruction is that a memory access was made, not where it was made. On the other hand, knowing whether the stack was accessed does reveal information about the executed instruction type because some instructions only operate on the stack and not on other segments of memory.

Therefore, instead of using the raw numbers contained in \instrmeas{}s, we collect four features: the execution latency (with a resolution of 10 cycles), the type of memory access (read/write or no access), whether the instruction accessed the stack (yes or no), and whether the instruction modified the control-flow (yes or no). %
We arbitrarily choose a 10-cycle resolution for the attacker to over-approximate the best current attacker capabilities. To the best of our knowledge, even the most advanced attacks that leverage instruction timings show significant noise and are not even close to a resolution of 10 cycles for current TEEs~\cite{frontal,nemesis}. More details on related attacks can be found in \cref{sec:relwork}.%
Note also that the \instrmeas does not include cache access information, despite being within the capabilities of a SotA today. We decided to exclude this information from the \instrmeas because the relevant features from this measurement (whether memory was accessed) can already be inferred from the page monitoring controlled-channel, which is easier to measure and deterministic. 
This highlights the difference between recovering instructions compared to data: for data inference, precise memory accesses are important, while for instruction inference, we need to extract \textit{metadata} about the instruction. 

\paragraph{Candidate sets}
To be able to quantitatively compare code leakage, we introduce the notion of \candidate sets. The attacker forms a \candidate set for each instruction they are trying to recover. %
Let us assume that from the \instrmeas, the attacker can deduce that the underlying confidential x86 instruction made a memory read from the stack, e.g., because the \instrmeas contains a memory read from a page assigned to the application's stack. Then the \candidate set for that \instrmeas will contain instructions such as \texttt{pop}, \texttt{mov}, and \texttt{add}, as they can all read from the stack. On the other hand, it will not contain a \texttt{push}, as this instruction always \emph{writes} to the stack. More formally, an instruction belongs to the \candidate set of an \instrmeas if and only if there exists a version of that instruction that would produce a set of observations that is exactly the \instrmeas. The \candidate set is useful in that it tells us that the instruction underlying an \instrmeas can only be among the ones contained in that \instrmeas \candidate set. Therefore, if the set only contains one instruction, then the attacker has recovered a target instruction. In general, we can say that the smaller the \candidate sets, the more information the attacker collected (i.e., the lower the entropy). The \candidate set allows us to compare the leakage in the two systems in the sense that if one system tends to produce smaller \candidate sets than the other, then we can say that it is leakier and by how much. The ISA used in the target system (x86 or WASM) helps in forming an initial \candidate set. Since the target system can only execute valid instructions, the candidate set of an instruction with an ``empty'' \instrmeas contains all of the instructions of the system's ISA. 

Finally, unless otherwise specified, we only report numbers for \emph{semantically different} instructions in the \candidate sets. We define semantic equivalent instructions are instructions that perform the same task but differ only in the input operand size or type (e.g., signed or unsigned). For instance, in WASM, \texttt{i32.add} is equivalent to \texttt{i64.add}, while in x86, \texttt{movq} is equivalent to \texttt{mov}. Semantically different instructions are then instructions that are not semantically equivalent. We perform this simplification because we note that generally, if a \candidate set contains only semantically equivalent instructions, it can be misleading to report a higher number of instructions in it.

\section{Leakage Analysis}\label{sec:leakage_analysis}

\begin{figure*}[!t]
\begin{minipage}[!t]{0.475\linewidth}
    \asmmintedfile{figures/small_loop.S}
    \caption{A simple assembly program with a loop that on each iteration computes $z = z * x$. The loop iterates $y$ times. The variable $x$ is stored on \texttt{\%eax}, y on \texttt{-8(\%rbp)}, and $z$ on \texttt{\%ecx}.}\label{fig:small_loop}%
\end{minipage}\hfill
\begin{minipage}[!t]{0.485\linewidth}
    \centering
    \captionof{table}{View of the attacker for the \texttt{asm} in \cref{fig:small_loop}; $y = 2$. Candidate sets contain only semantically different instructions. Collected in the Skylake microarchitecture.}%
    \label{tb:att_view}
    \resizebox{\linewidth}{!}{%
        \begin{NiceTabular}{@{}lcccc|c@{}}
        \CodeBefore
            \rowcolors{3}{blue!3}{}
        \Body
        \toprule
        Instruction       & Cycles         & Memory        & Stack Access & Is CF? & \Candidate set size   \\ \midrule
        \texttt{mov}      & $0 - 10$       & R             & \Yes         & \No    & 545                   \\
        \texttt{imul}     & $0 - 10$       & -             & \No          & \No    & 581                   \\
        \texttt{mov}      & $0 - 10$       & W             & \Yes         & \No    & 86                    \\
        \texttt{inc}      & $0 - 10$       & -             & \No          & \No    & 581                   \\
        \texttt{cmp}      & $0 - 10$       & R             & \Yes         & \No    & 545                   \\
        \texttt{jmp}      & $0 - 10$       & -             & \No          & \Yes   & 23                    \\ \hline
        \texttt{mov}      & $0 - 10$       & R             & \Yes         & \No    & 545                   \\
        \texttt{imul}     & $0 - 10$       & -             & \No          & \No    & 581                   \\
        \texttt{mov}      & $0 - 10$       & W             & \Yes         & \No    & 86                    \\
        \texttt{inc}      & $0 - 10$       & -             & \No          & \No    & 581                   \\
        \texttt{cmp}      & $0 - 10$       & R             & \Yes         & \No    & 545                   \\
        \texttt{jmp}      & $0 - 10$       & -             & \No          & \Yes   & 23                    \\
        \bottomrule                
        \end{NiceTabular}%
        }

\end{minipage}\hfill
\end{figure*}

We now explain how to leverage \instrmeass to build \candidate sets for instructions in the native and WASM systems and use such \candidate sets to measure how much of the confidential code leaks.
For both systems, we proceed as follows:
\begin{itemize}
    \item First, we analyze a simple program: a small loop where each iteration computes the multiplicative product of two numbers, reported in \cref{fig:small_loop}. It is composed of $6$ assembly instructions, where the two numbers are multiplied in line $4$. We compile this program to x86 for the \baseline and to WASM for the \interpreted.
    \item Second, we discuss the \instrmeass obtained from its execution and analyze the candidate set sizes for each instruction.
    \item Finally, we estimate the leakage of the system by computing candidate set sizes for all instructions in its ISA.
\end{itemize}
In the following, we first analyze the baseline native system. We start our analysis with the SotA attacker with practical capabilities (e.g., timing resolution of 10 cycles). We then expand the attacker capabilities to account for future attacks with single-cycle accuracy, functional units occupied over time, and more. We use such an unrealistically strong attacker to determine an upper bound to leakage in the native system (\cref{sec:ideal}). 
Finally, we analyze the WASM system under the SotA attacker (\cref{sec:wasm_study}).

\subsection{Leakage in the \BaselinE}\label{sec:baseline}

We compiled the sample binary from WASM bytecode to x86 and profiled its execution to gather its \instrmeass: \Cref{tb:att_view} shows the collected features when the loop is executed twice and the number of \emph{\candidate instructions} on Skylake CPUs. We observe that, despite combining the information from several side-channels, rarely the attacker gets a \candidate set with fewer than $100$ instructions. %

In fact, this is not the case just in the example binary of \cref{fig:small_loop}, but it is a consequence of the classes of instructions that can be built with the employed side-channels. As there are fewer classes than there are instructions, some instructions are bound to belong to the same \candidate set, thus making them indistinguishable from each other.

\paragraph{Full ISA}
We now turn to the full native system ISA: by computing all possible \candidate sets, we can check how many instructions of the ISA have a \candidate set size below a certain threshold, with the idea that the smaller the overall \candidate set sizes are, the leakier a system is. Observe that each \instrmeas maps to exactly one class, and the instructions in that class form the \candidate set for that \instrmeas. This means that all the possible classes are exactly all the possible \candidate sets that can be observed for a system. 

However, to estimate which instructions are in which class, we would need to collect an \instrmeas for all instructions (and their variations) in the x86 ISA available in SGX and SEV. Further, we would also have to do this for different microarchitectures, as these support different extensions of the x86 ISA and thus change the set of available instructions. Instead of generating programs to execute all possible instructions on different microarchitectures, we adapted and reused the results of a dataset collected as part of an x86 benchmarking suite for the x86 ISA~\cite{abel19a}. Particularly, we inferred from the dataset to which class among the ones introduced above every instruction belongs.
The dataset had to be adapted to account for the fact that some instructions are illegal in SGX or that others are intercepted by the hypervisor on SEV. We describe these caveats in \cref{sec:dataset_desc_sgx}.

\begin{figure}[tbp]
    \centering
    \includegraphics[trim={0 0 0 0},clip,width=\linewidth]{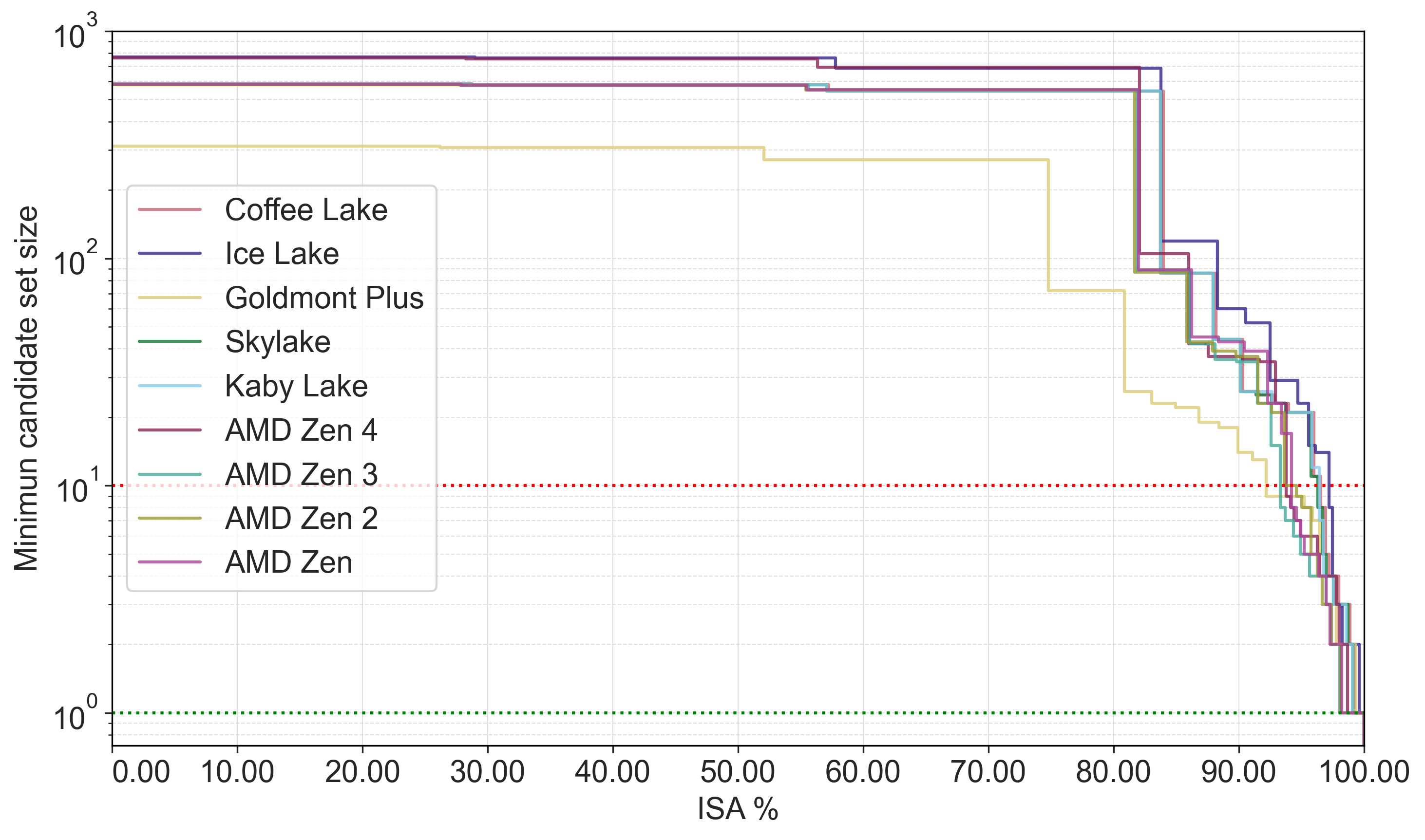}
    \vspace{-15pt}
    \caption{Instruction \candidate set size distribution of semantically different SGX and SEV instructions for various 64-bit x86 microarchitectures under the SotA attacker. The plot shows the minimum \candidate set size that contains at least $x$ percent of the ISA available in the TEE (SEV for AMD and SGX for Intel). This assumes the best resolution available to the SotA attacker with respect to execution time is $10$ cycles. The dotted red line is set at $y=10$, and it indicates that $>90\%$ of the ISA instructions have a \candidate set size greater than $10$.} %
    \label{fig:x86_instr_distr}
\end{figure}

We report the cumulative distribution of the sizes of the \candidate sets in \cref{fig:x86_instr_distr}. %
What can be observed from the figure is that around 80\% of the instructions of the ISA belong to a \candidate set containing more than $100$ instructions. Note that for SEV, 1.48\% of the instructions in the ISA belong to a \candidate set of size $1$ and can therefore be leaked to the attacker. This is due to the fact that in SEV, some instructions, such as \texttt{CPUID}, are intercepted by the hypervisor and are therefore leaked to the attacker (not through side-channels, but through a system interface). 
There are a few other instructions with a candidate set size of $<10$, but they are limited to less than 8\% for all analyzed microarchitectures.
Thus the SotA attacker is practically never able to resolve any instruction of the x86 ISA \emph{based on the side-channel information alone}. %

\subsection{Ideal Attacker}\label{sec:ideal}

For the \baseline, we also explore different strengths of attacker models, for instance, showing how the candidate set sizes change based on different levels of cycle accuracy available to the attacker. We present these results in \cref{sec:sota_thresholds} and discuss in \cref{sec:relwork} how these resolutions map to known attacks. Here instead, we study what we believe to be the extreme in terms of attacker strength, which we refer to as the \emph{ideal} attacker. %
The ideal attacker has the capability of benchmarking instructions, as done in~\cite{abel19a}.
Note that~\cite{abel19a} is a general method to benchmark instructions outside the enclave and hence uses capabilities currently blocked by SGX and SEV, such as reading performance counters and injecting instructions around target instructions. 
These capabilities also allow the attacker to observe the utilization of individual functional units and obtain cycle-accurate execution time for each instruction. 
We assume that the other security properties of SGX and SEV otherwise hold, e.g., the ideal attacker cannot read the enclave memory. 
To the best of our knowledge, the data on single instructions collected in \cite{abel19a} is the most detailed and comprehensive dataset about the performance of current x86 processors to date. Since current attacks do not even get close to the resolution and wealth of information available in \cite{abel19a}, the ideal attacker is currently far from realistic. Nonetheless, we see value in this second attacker model as it allows us to reason about a leakage model against a theoretically stronger attacker and to establish an upper bound of leakage that can be achieved.

\begin{figure}[t!]
    \centering
    \includegraphics[trim={0 0 0 0},clip,width=\linewidth]{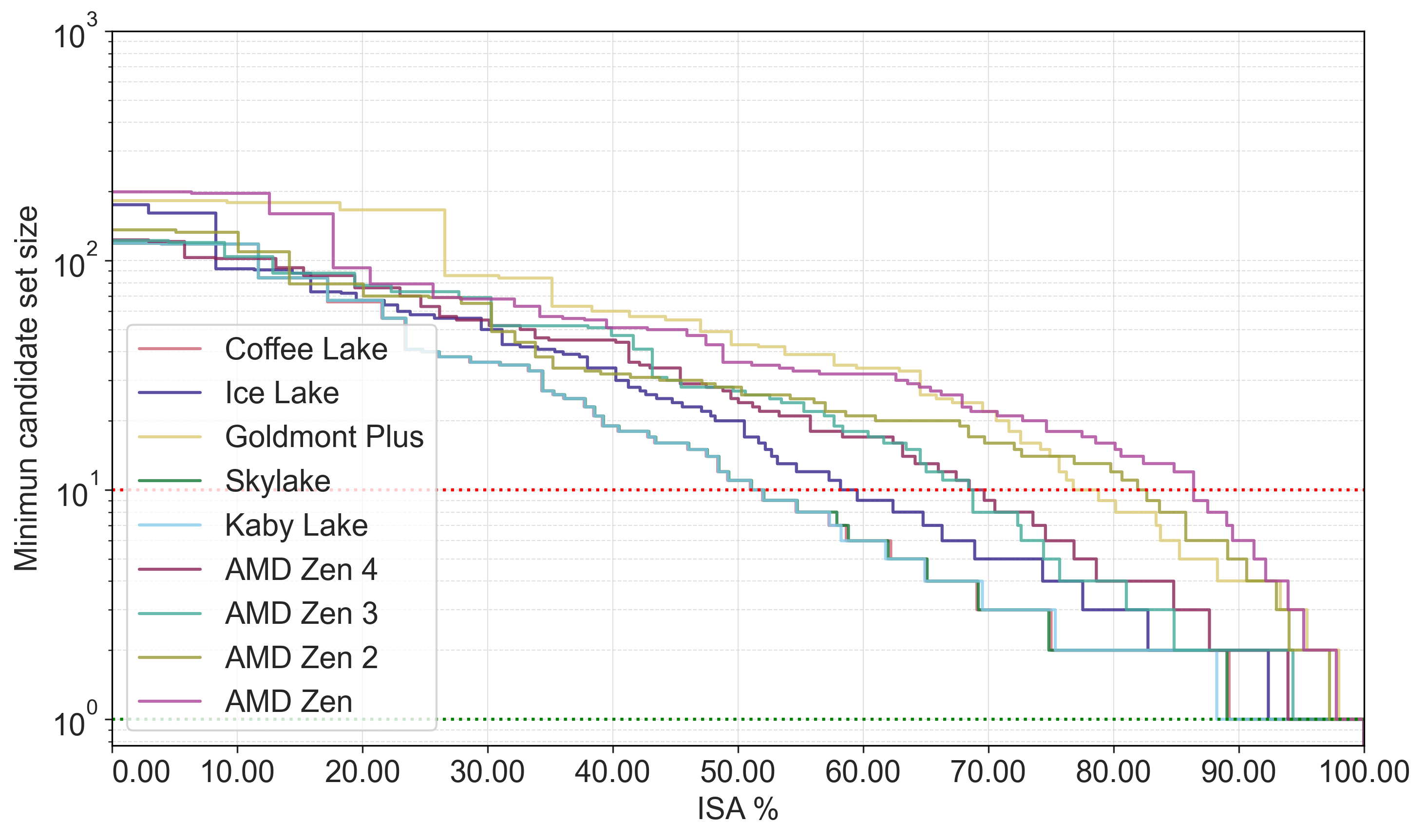}
    \vspace{-15pt}
    \caption{Instruction \candidate set size distribution of semantically different SGX and SEV instructions for various 64-bit x86 microarchitectures under the ideal attacker. The dotted green line is set at $y=1$, and given where it intersects the various microarchitectures' ISA, it indicates that more than 90\% of the instructions cannot be recovered even by the ideal attacker.} %
    \label{fig:ideal_instr_dist}
\end{figure}

To build the candidate sets for the ideal attacker, we construct the \instrmeas using the data in \cite{abel19a} as follows: cycle-accurate execution time, functional units (FUs) occupied over time, the code address accessed, the data address accessed (if any), and the type of data access (read or write). Regarding the FUs, for each instruction, we let the attacker perfectly see the order in which they are used and which other FUs could be used by the instruction. Using these very detailed \instrmeass, we create the candidate sets by grouping together all x86 instructions for which the information in the \instrmeas is exactly the same\footnote{Exclude code and data addresses as these only contain information related to the input data and not the confidential instruction.}. Finally, we remove duplicate entries that are semantically similar, e.g., \texttt{mov} and \texttt{movq}. The resulting cumulative distribution of the \candidate set sizes is depicted in \cref{fig:ideal_instr_dist}.

While the resulting candidate set sizes are significantly smaller than for the SotA attacker, around 50\% of the ISA still belongs to a candidate set of at least size $10$ for all analyzed microarchitectures. On the other hand, up to 10\% of instructions are uniquely identifiable with a candidate set of size $1$ on both SGX and SEV. Based on these results, the ideal attacker might be able to extract some instructions, but the majority of the ISA still remains ambiguous and cannot easily be leaked. Therefore, even an unrealistically strong adversary is not able to reconstruct most confidential x86 instructions from the \instrmeass{} side-channels.

\subsection{Leakage in the \interpreteD}\label{sec:wasm_study}

We again first consider the loop of \cref{fig:small_loop}, compiled to WASM.
However, while in the native system the binary only gets executed, we note that WASM translators (AOT, JIT, and pure interpreters) generally have two phases: \emph{loading} and \emph{interpretation}. During loading, the WASM binary is parsed, and each instruction is decoded into some internal and implementation-specific format. The second phase encompasses the execution of the loaded WASM binary. 

We choose to analyze the WAMR~\cite{wamr} interpreter because it combines aspects of both a JIT compiler and a pure interpreter. During the loading phase, WAMR parses the WASM instructions and eliminates instructions whose results can be statically determined. 
For instance, the loader optimizes away instructions that load constant parameters by pre-placing their constants into the WASM stack before execution.
This optimization speeds up the interpreter, as only a subset of instructions needs to be executed later. This pre-processing of instructions makes the loading phase of WAMR akin to a JIT compiler.
Multiple native x86 instructions are executed for each WASM instruction during both phases. Thus, each WASM instruction of the loop of the sample program lets us collect multiple \instrmeass: we report them in \cref{tb:wasm_att_view}. In WASM, the loop is composed of $20$ instructions, out of which $12$ are simplified in the loading phase, leaving $8$ instructions (marked in bold in the table) to be executed in the interpreter phase. 
In total, we recorded $1290$ \instrmeass in the loading phase of the loop and $184$ \instrmeass in the interpreter phase (with two loop iterations). Between loading and interpreting the loop, the WASM system presents a $123$x increase in instructions executed compared to when the same code is executed in the native system.

\begin{table}[t]
    \caption{Attacker view of the loop in \cref{fig:small_loop} in the WAMR loader and interpreter. Normal font instructions are optimized away by the JIT loader, while bold instructions are executed also when interpreting. We use the same version of WAMR as in \cref{lst:wasm_loader}. We give only one loop iteration - only bold instructions repeat on each iteration.} 
    \label{tb:wasm_att_view}
    \centering
    \resizebox{\linewidth}{!}{%
        \begin{NiceTabular}{@{}lcc|cc@{}}
        \CodeBefore
            \rowcolors{3}{}{blue!3}
        \Body
        \toprule 
        \multirow{2.4}{*}{Instruction} & \multicolumn{2}{c}{\# of \instrmeas per instruction} & \multicolumn{2}{c}{\Candidate set size} \\ \cmidrule(lr){2-3}\cmidrule(lr){4-5}
                                       & JIT Loader & \multicolumn{1}{c}{Interpreter}         & JIT Loader & Interpreter \\\midrule
        \texttt{loop}                  & 66         & \multicolumn{1}{c}{-}                   & 1          & -           \\
        \texttt{get.local}             & 63         & \multicolumn{1}{c}{-}                   & 1          & -           \\
        \texttt{get.local}             & 62         & \multicolumn{1}{c}{-}                   & 1          & -           \\
        \texttt{get.local}             & 63         & \multicolumn{1}{c}{-}                   & 1          & -           \\
        \textbf{\texttt{i32.mul}}      & 33         & \multicolumn{1}{c}{9}                   & 4          & 6           \\
        \textbf{\texttt{i32.store}}    & 91         & \multicolumn{1}{c}{14}                  & 1          & 1           \\
        \texttt{get.local}             & 63         & \multicolumn{1}{c}{-}                   & 1          & -           \\
        \textbf{\texttt{i32.load}}     & 91         & \multicolumn{1}{c}{14}                  & 1          & 1           \\
        \texttt{set.local}             & 80         & \multicolumn{1}{c}{-}                   & 1          & -           \\
        \texttt{get.local}             & 63         & \multicolumn{1}{c}{-}                   & 1          & -           \\
        \textbf{\texttt{i32.load}}     & 91         & \multicolumn{1}{c}{14}                  & 1          & 1           \\
        \texttt{set.local}             & 80         & \multicolumn{1}{c}{-}                   & 1          & -           \\
        \texttt{get.local}             & 62         & \multicolumn{1}{c}{-}                   & 1          & -           \\
        \texttt{i32.const}             & 55         & \multicolumn{1}{c}{-}                   & 1          & -           \\
        \textbf{\texttt{i32.add}}      & 33         & \multicolumn{1}{c}{9}                   & 4          & 6           \\
        \textbf{\texttt{local.tee}}    & 97         & \multicolumn{1}{c}{7}                   & 1          & 2           \\ 
        \texttt{get.local}             & 62         & \multicolumn{1}{c}{-}                   & 1          & -           \\
        \textbf{\texttt{i32.lt\_s}}    & 33         & \multicolumn{1}{c}{11}                  & 4          & 6           \\
        \textbf{\texttt{br\_if}}       & 35         & \multicolumn{1}{c}{14}                  & 1          & 1           \\
        \texttt{end}                   & 67         & \multicolumn{1}{c}{-}                   & 1          & -           \\
        \bottomrule                         
        \end{NiceTabular}
        }
\end{table}

Our goal is now to understand how \textit{unique} each trace of \instrmeass for each of these WASM instructions is.
For this, we profiled each WASM instruction (see \cref{sec:impl} for more details) and obtained their traces of \instrmeass.
With this profiling, we build \candidate sets for the WASM system using the information obtained from multiple \instrmeass to differentiate instructions. \Cref{tb:wasm_att_view} reports the \candidate set sizes for the instructions in the loop. For several instructions, the attacker gets \candidate set sizes of size 1, thus perfectly recovering the instruction, which was not possible in the \baseline. 

However, even in WASM, some instructions are very similar to each other, e.g., instructions that require few x86 instructions to execute tend to still be challenging to classify accurately. For instance, in the WAMR interpreter, the \texttt{i32.add} and \texttt{i32.sub} instructions are both implemented with $9$ x86 instructions and differ for a single one: \texttt{i32.add} uses an x86 \texttt{add} where \texttt{i32.sub} has an x86 \texttt{sub}. 
Since the side-channels available to the SotA attacker cannot distinguish between these two instructions, \texttt{i32.add} and \texttt{i32.sub} end up in the same \candidate set\footnote{Interestingly, the attacker can still distinguish these two instructions because they differ in multiple instructions in the loading phase.}. We can also observe this in \cref{tb:wasm_att_view}: instructions with a small number of \instrmeass tend to have bigger \candidate set sizes.

\begin{figure}[t!]
    \centering
    \includegraphics[trim={0 0 0 0},clip,width=\linewidth]{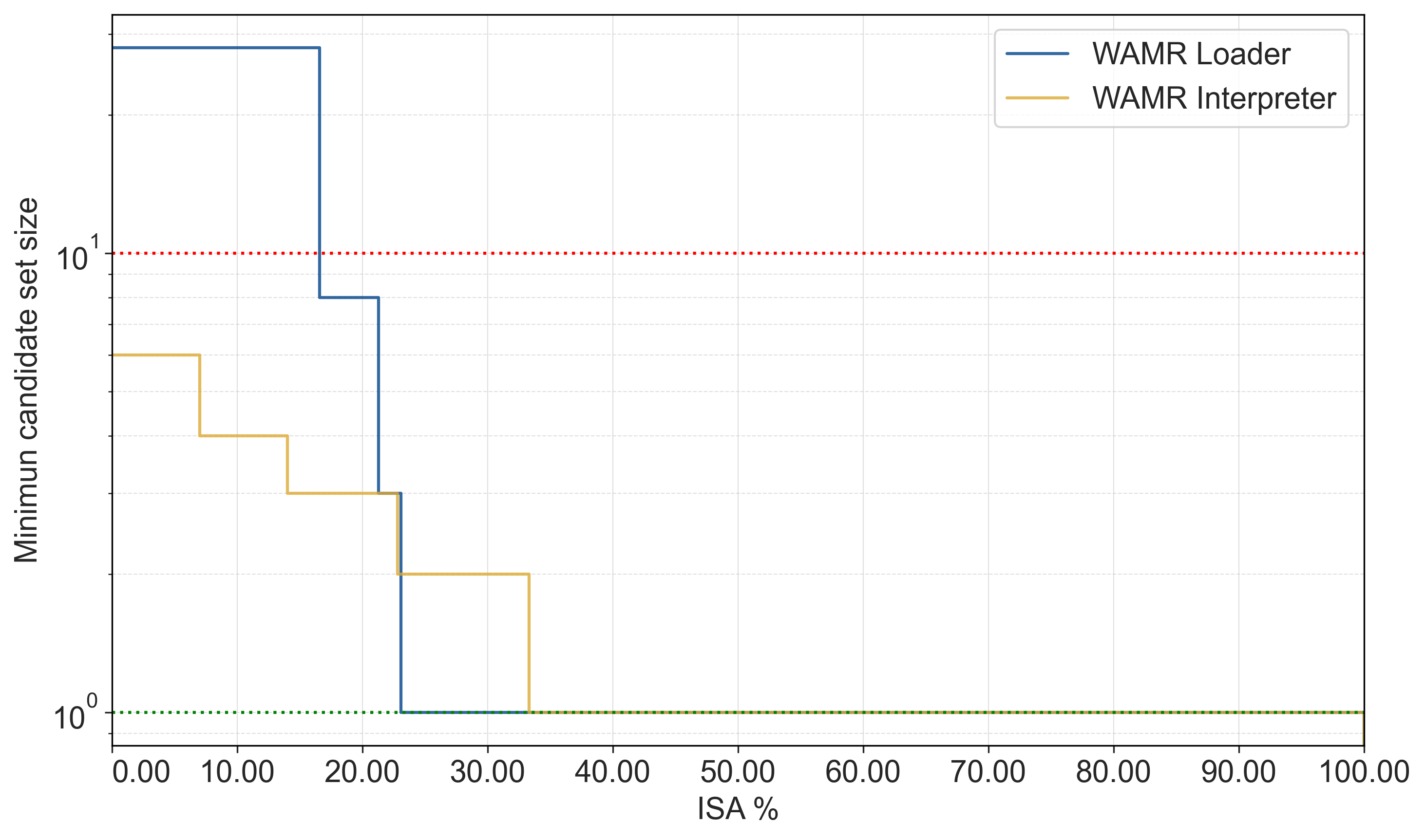}
    \vspace{-15pt}
    \caption{\Candidate set size distribution of WASM instructions in the WAMR interpreter under the SotA attacker. Only semantically different instructions are included in the \candidate sets. The green dotted line is at $y = 1$, where \candidate sets of that size offer no confidentiality.}
    \label{fig:wasm_instr_dist}
\end{figure}

In summary, the WASM system leaks more instructions of the example loop compared to the native system, with 85\% of its instructions being fully leaked (compared to 0\% in the \baseline). 

\paragraph{Full ISA}
Similarly to the native system, we compute all possible \candidate sets of the WASM system: if the \candidate sets tend to be small for a large percentage of the WASM ISA, then the system itself cannot provide code confidentiality, as this attack will likely extend to different WASM binaries besides our sample program.

For this, we obtained the \instrmeass of each WASM instruction while profiling a WASM test suite~\cite{webassembly_test_suite}. The test suite we used is developed to comply with the WASM standard and ensures we reach a good coverage for all the $172$ core WASM instructions. 
We depict the distribution of the WASM instructions' \candidate sets that we obtained for WAMR in \cref{fig:wasm_instr_dist}. As can be seen, almost 80\% of the ISA has a \candidate set size $\leq 2$, both in the loading phase and the interpreter phase (which can be combined in a real attack).
Compare this to the \baseline, where even the ideal attacker could, at best, recover 10\% of the ISA instructions, and it is clear that the WASM system is leakier than the \baseline. Finally, not only is the WASM system leakier, but the results also highlight that a SotA attacker can practically break code confidentiality for at least 70\% of the WASM ISA. %

\section{IR Instruction Leakage in Practice}
\label{sec:impl}

\begin{figure*}[t]
    \centering
    \includegraphics[trim={0 0 0 0},clip,width=\linewidth]{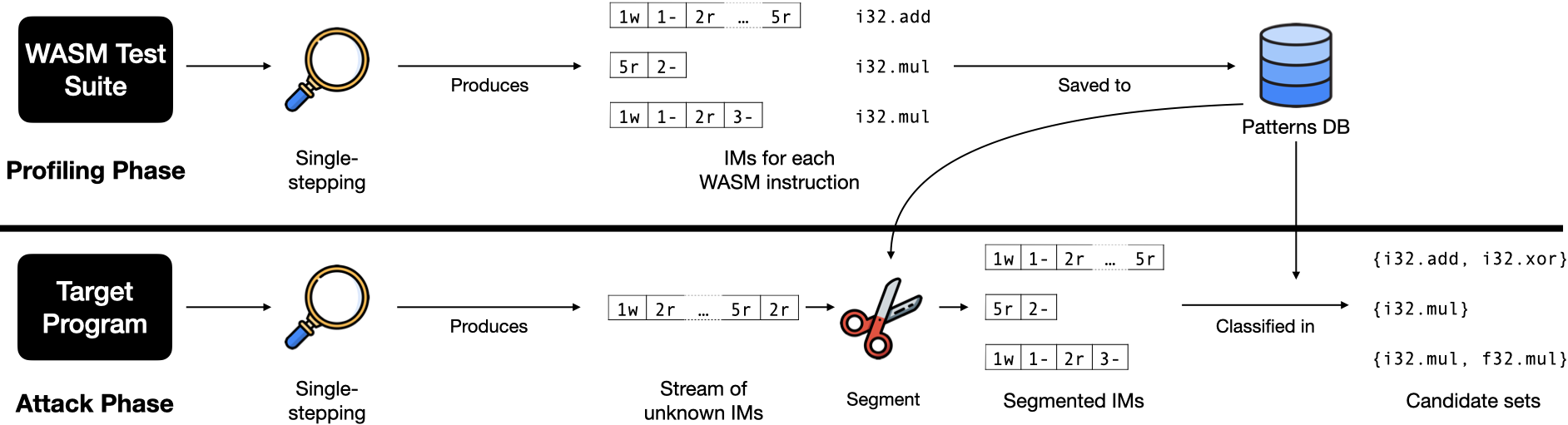}
    \vspace{-15pt}
    \caption{Overview of the end-to-end attack steps.}
    \label{fig:wasm-atk-overview}
\end{figure*}

We now describe how to extract confidential WASM instructions from the WASM system. We focus on SGX due to the availability of better tooling in this platform. We discuss in \cref{sec:discussion} to what extent these results extend to SEV as well.

We depict our attack in \cref{fig:wasm-atk-overview}.
In the \textbf{Profiling Phase}, the attacker single-steps the enclave execution to collect \instrmeass{} for each possible WASM instruction and generates a database of patterns.
We detail this phase in \cref{sec:impl-profiling}.
In the \textbf{Attack Phase}, the attacker again single-steps the enclave execution while the target WASM program is being interpreted.
Here, the attacker obtains a single stream of \instrmeass that need to be segmented correctly before matching each segment with the previously profiled patterns.
We describe this phase in \cref{sec:impl-attack}.

\subsection{Profiling Phase}
\label{sec:impl-profiling}
\label{sec:patt_generation}

In this phase, the attacker's goal is to profile the target translator and generate patterns of traces for each WASM instruction.
To do so, the attacker follows the methodology described in \cref{sec:methodology}: having complete control over the enclave during this phase, the attacker can know exactly which \instrmeass correspond to which WASM instruction. 
For example, we do so by saving the Instruction Pointer (IP) together with the measurements and obtaining the ground truth of the parsed instructions from a modified translator\footnote{The instruction pointer is not available to the attacker in the attack phase but represents valuable information: printing such ground truth of executed instructions is helpful for verification purposes, as the translator might, e.g., parse some instructions twice or skip some of them.}.

To build an extensive dataset for the translator, the attacker needs to profile a program that calls as many WASM instructions as possible, feeding different input data to reach good coverage\footnote{Feeding different input values is important because the same WASM instruction might be executed by a different set of x86 instructions depending on what inputs are given to it, as we discuss in \cref{sec:regex_gen}.}.
We use the official WASM test suite~\cite{webassembly_test_suite}, maintained by the WebAssembly Working Group that is used to test the adherence of new compilers and interpreters to the WASM specification.

On the translators that we tested, we empirically verified that we only need two pieces of information for every \instrmeas to segment an execution trace: the code page number that was accessed and whether the x86 instruction performed a memory read, a memory write, or no memory access. Thus we represent each \instrmeas with a string composed of two parts: (i) the code page number; and (ii) the memory access type, e.g., \texttt{1r} represents an x86 instruction that was executed from page number $1$ and made a memory read. Similarly, \texttt{1w} and \texttt{1-} refer to a memory write and to no memory access, respectively, from an instruction executed on page $1$. When a WASM instruction is composed of multiple x86 instructions, we concatenate these symbols for the various \instrmeass that were recorded for that WASM instruction. We refer to this string as the \textit{pattern} for a particular WASM instruction.

\subsubsection{Profiling the WAMR interpreter}
We now further discuss how we extract these patterns in the WAMR~\cite{wamr} interpreter during both of its execution phases: loading and interpreting.

\paragraph{Loading}
During the loading phase, WAMR loops through each instruction, as shown in \cref{lst:wasm_loader}.
By manually inspecting the binary of the WAMR interpreter, we found the addresses of the first instruction of this loop and the first instruction outside of the loop. 
Segmenting the loader execution trace is straightforward with knowledge of the IP: we look for the loop's entry point and create a new segment every time the entry point's IP is found in the instruction trace. When we encounter the first instruction outside of the loop, we stop segment generation and restart it when we encounter the beginning of the loop again. 
This gives us a pattern for each loop iteration: to know which WASM instruction corresponds to each iteration we modify the WAMR interpreter to record which instruction was parsed in which iteration.
Note here that \textit{only} the ground truth is obtained from a modified WAMR version: the execution trace to attack is obtained from an unmodified version.

\paragraph{Interpreting}
In the interpreter phase, the WAMR control-flow is more involved than in the loading phase. The interpreter executes one instruction, then fetches the pointer of the next instruction from memory and directly jumps to it - without any loop.
Crucially, every jump to the next WASM instruction is implemented as an indirect jump (e.g., \texttt{jmp *rax}). Thus, to segment the execution trace of the interpreter, we look for indirect jumps in the execution trace. Since we have the IP for each \instrmeas, we can check on the WAMR interpreter whether the instruction at that IP is an indirect jump. Whenever we encounter an indirect jump, we create a new segment\footnote{This approach only works if indirect jumps are used only at the boundary between two instructions, as is the case in the WAMR interpreter.}. 
Similarly to the loader, we need to label the segments: we again modified the WAMR interpreter to print the instruction label every time it starts interpreting a new instruction to get the ground truth of labels for each test in the test suite. We then assign the labels to each of the segments obtained by monitoring the IP of the execution trace.

The process would follow a similar flow in other translators: what the attacker needs is a way to find instructions boundaries based on the IP (either by manual inspection or automatically) and a way to map each segment to (known) WASM instructions.

\subsubsection{Fused instructions handling}
The way we build patterns for WASM instructions does not properly account for \textit{fused instructions} from the CPU: separate x86 instructions that the CPU executes as one. 
When single-stepping with interrupts, these instructions will be stepped through atomically -- thus, we will only encounter one \instrmeas in the execution trace instead of two. This phenomenon has been documented in previous work as well~\cite{frontal, copycat}. However, while previous work observed deterministic instruction fusion~\cite{copycat}, we observed that for the same pair of instructions in the program (at the same virtual address), it can happen that the instructions sometimes execute unfused. This is the case even when the same input data is given to the program both with and without hyperthreading enabled. We hypothesize that this behavior is due to the precise timing at which the interrupt is delivered in relation to the stage of the execution of the to-be-fused instruction pair. However, the timing at which the interrupt is delivered cannot be controlled to such precision, and therefore the behavior randomly occurs, albeit somewhat infrequently. Note that we collect significantly larger instruction traces compared to~\cite{copycat} (e.g., more than 1 billion instructions) and hence have a higher likelihood of observing this behavior compared to~\cite{copycat}.

Unfortunately, this leads to the pattern of WASM instructions being non-deterministic. Theoretically, we could collect every possible variation of one WASM instruction, repeating a trace collection many times until we get all possible patterns. However, this approach is infeasible in practice for two reasons. First, since the CPU non-deterministically fuses instructions, collecting all possible patterns for a WASM instruction requires a lot of repetitions and is not guaranteed to terminate. Second, the number of different traces needed to be collected grows exponentially with the number of possible fused instruction pairs. Just having $10$ fused instructions pairs in a trace requires $1024$ patterns to be collected and stored.

We addressed this issue by detecting which \instrmeas could be related to fused instructions and then saving only the fused version of the pattern. Alongside the pattern, we save an array of positions that could be potentially ``unfused''. This representation is not only compact (we need to save only one version of the pattern) but also allows us to efficiently match any combination of unfused instructions in the pattern. Knowing which \instrmeass are related to fused or unfused instructions is done by cross-referencing the x86 instructions of the WAMR loader with the IP recorded for the \instrmeas.

\subsection{Attack Phase}
\label{sec:impl-attack}

\subsubsection{Trace segmentation}\label{sec:segmentation}
In the attack phase, the adversary now targets a production enclave with the target confidential algorithm and profiles it to obtain an execution trace.
As the attacker cannot obtain the IP, segmenting the different \instrmeass for each WASM instruction is more difficult in this phase.
However, by representing the full execution trace as a string, we can reduce it to the well-known string-matching problem. 
Segmenting the trace then proceeds as follows: starting from the beginning of the string, we try to match all of the previously collected patterns. We then take one of the matches\footnote{Multiple matches are possible because patterns overlap.} and advance the starting pointer to just after these instructions. We then try to match a new pattern to this position in the string. If nothing matches, we backtrack and choose one of the previously found valid patterns. We repeat this process until the whole execution trace is perfectly segmented. We will discuss the performance of this algorithm in practice in \cref{sec:eval}.

\subsubsection{Creating and matching patterns}\label{sec:regex_gen}
The approach described above assumes that we can collect \emph{every} possible pattern for each WASM instruction. 
Whilst the test suite patterns achieve a wide coverage, we still do not collect enough patterns to fully segment unseen binaries.
In particular, while \textit{linear} WASM instructions (WASM instructions that have no loops or branching conditions) exhibit only a single pattern, it is challenging to build every pattern for instructions with loops and branches. For instance, the WASM \texttt{clz} instruction is implemented in the interpreter with a loop that iterates once for every leading zero present in the input integer. 

For cases of instructions with complex control flow, we leverage the observation that generally, their start instructions and end instructions will be the same, no matter how complex the internal control-flow is. Thus when we encounter more than one pattern for the same instruction, we automatically try to generalize its pattern. We do this by arranging the characters of the string representation in a tree where each node of the tree is one \textit{token} (code page number and access type). We then add multiple patterns to the same tree and extract the common prefixes from it. Particularly, after the tree is assembled, we traverse it and collect every pattern found up to 2-3 splits of the tree. We found this heuristic to be quite accurate in practice. We do the same process both to find common beginning prefixes and end suffixes.

Between matching for common prefixes and suffixes and accounting for variable numbers of instructions due to fused instructions, we found that the most convenient way to apply the patterns was through regular expressions (\textit{regexes}). This allowed us to use already existing and optimized matching engines and rapid prototyping of different matching configurations. We automatically generated regexes for each possible segment while also keeping the regexes' complexity within bounds.

\begin{figure}
    \centering
    \includegraphics[width=1\linewidth]{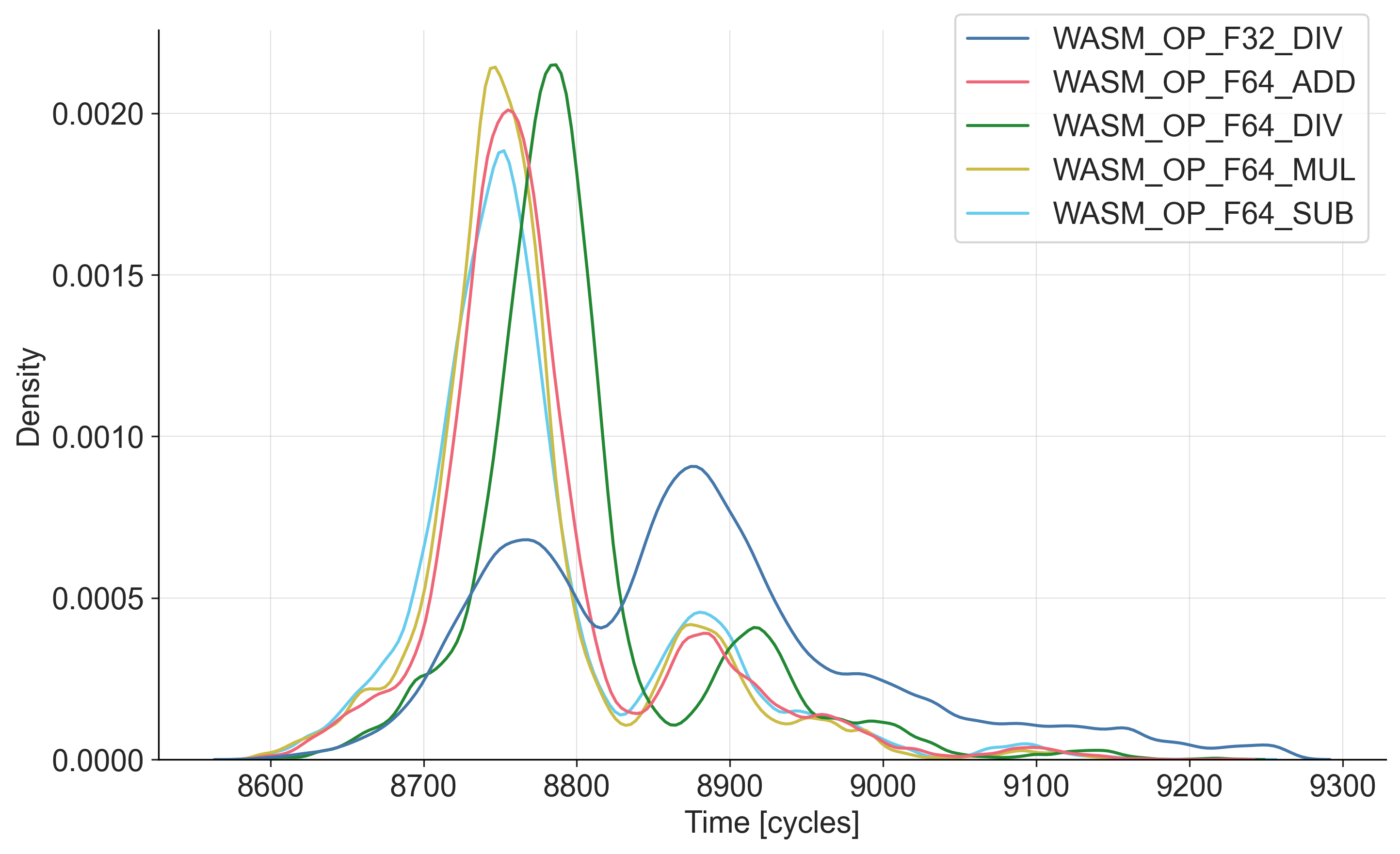}
    \vspace{-15pt}
    \caption{Timing distribution of the 5th x86 instruction for the five listed WASM instructions. The two division operations seem to be following a different distribution than the others. N=11527}
    \label{fig:timingdist}
\end{figure}

\subsubsection{Segment classification}
As discussed above, trace segmentation and segment classification are inherently linked tasks. Given a correct segmentation, we already get ``for free'' a possible list of candidate WASM instructions for each segment: those are the instructions whose known patterns matched the segment.
We call this a \textit{candidate set}.
In fact, this is how we generated the \candidate sets for WASM that we discussed in \cref{sec:wasm_study}. Recall that segments are generated only using the code pages and the memory access type: this information alone is so accurate to not only segment the trace but also to perfectly classify up to 80\% of the WASM ISA.

\subsubsection{Candidate sets pruning}
We investigate whether we can further reduce the candidate set size for the remaining 20\% of the WASM ISA where there is more than one candidate.
In particular, \instrmeass also contain the time spent executing individual x86 instructions, a feature that we did not use so far in our attack, as it is not fully deterministic. %
We explore the potential of using time measurements to prune candidate sets with a concrete \candidate set obtained from the WAMR interpreter containing the following WASM instructions: \texttt{F32\_DIV}, \texttt{F64\_ADD}, \texttt{F64\_DIV}, \texttt{F64\_MUL}, and \texttt{F64\_SUB}. Manual inspection of the interpreter's binary reveals that all of these WASM instructions are expanded into $9$ x86 instructions. However, among these $9$, only the fifth x86 instruction differs between the WASM instructions. %
Therefore, any potential timing difference should be visible only in the 5th instruction\footnote{We observed that surrounding instructions are also affected and exhibit timing differences, albeit smaller ones.}. The distribution of the recorded timings of the 5th x86 instruction is depicted in \cref{fig:timingdist}. While the timing distributions mostly overlap, they still exhibit some differences between them.

\begin{figure}
    \centering
    \includegraphics[width=0.8\linewidth]{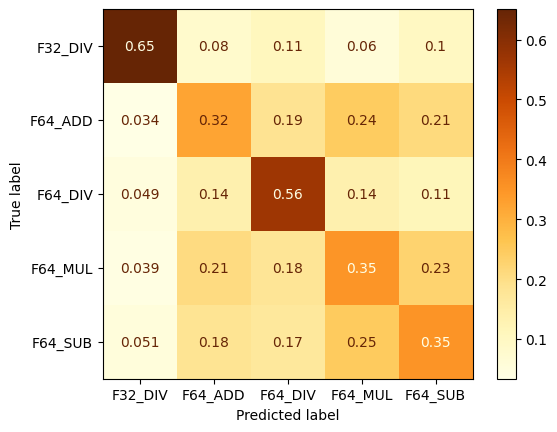}
    \vspace{-10pt}
    \caption{Confusion matrix of a simple random forest classifier for five WASM instructions. The classifier is pretty confident about the two divisions but cannot distinguish between the other three instructions.}
    \label{fig:confusion}
\end{figure}

To demonstrate the significance of these timing differences, we developed a very simple classical machine learning model that tries to classify between the aforementioned five  WASM instructions using only the timing data. A simple random-forest classifier~\cite{ho1995random} achieves around 45\% accuracy, significantly outperforming a random guess (which has 20\% accuracy). A confusion matrix is shown in \cref{fig:confusion}. 

In summary, the \candidate sets that we presented in \cref{fig:wasm_instr_dist} could be improved by including timing information. However, the attacker would have to record multiple executions for the same confidential algorithm to establish some confidence in the results. On the other hand, the information used when segmenting is deterministic, so the attacker only needs one execution of the confidential code to build the \candidate sets that were presented in \cref{fig:wasm_instr_dist}, and we thus deem the deterministic pipeline to be sufficient in practice.

\section{Evaluation}\label{sec:eval}
We evaluated the methods and algorithms presented in \cref{sec:impl} by using an Intel SGX enclave running the WAMR~\cite{wamr} runtime at commit version \texttt{b554a9d}. To collect the patterns for each WASM instruction, we single-stepped WAMR while it was executing the WASM test suite~\cite{webassembly_test_suite} (commit  \texttt{e87021b}). We run only tests that do not test for exceptions, as we are interested only in correct programs, although it would be straightforward to also include these tests.

\paragraph{Pattern generation} Overall, we profiled $21073$ tests. Note that we single-step the test suite with the enclave in debug mode, as we need the IP to produce the segmentation patterns as discussed in \cref{sec:patt_generation}. When we are profiling the WAMR loader, we only single-step the loader function (\texttt{wasm\_loader\_prepare\_bytecode}). When we are profiling the interpreter, we focus only on the interpreter's main function (\texttt{wasm\_interp\_call\_func\_bytecode}). By monitoring the program counter after the trace collection, we observed that we can very reliably single-step the enclave through interrupts, as no instruction was skipped for any of the tests in the test suite. Hence we run each test only once. In our machine (with an Intel i9-9900KS CPU), this takes about $24$ hours for the loader and about $36$ hours for the interpreter. In total, we found $1576$ unique patterns for the loader and $345$ unique patterns for the interpreter. Using the methods described in \cref{sec:segmentation}, we then created $137$ regular expressions for the loader patterns and $133$ for the interpreter ones. %

\paragraph{Instruction matching}
The SotA adversary can be instantiated in practice, and thus we performed our evaluation with real-world experiments. To test the generality and usefulness of the patterns, we used them to classify single WASM instructions in three synthetic programs. One of the programs is written in C and computes various cryptographic functions. The other two are written in Rust. One is part of a chess engine~\cite{rust_chess}, while the other computes the hash of its inputs. We compiled these programs to WASM and then gave them as input to an initial enclave running WAMR. We single-stepped this enclave in production mode (i.e., without getting the IP information). Not all possible instruction patterns of these programs were present in the test suite. We verified this by naively trying to match the patterns we collected from the test suite and found that some parts of the trace could not be segmented. However, we were able to fully segment the trace using generalized regex patterns. The C code, the Rust chess code, and the hash code executed $474 M$, $431 M$, and $62 k$ WASM instructions, respectively. When loading the code, they parsed $9 k$, $38 k$, and $49 k$ WASM instructions, respectively. Single stepping the interpreter phase took around $10$ hours for both the chess engine and the C code and a couple of seconds for the hash engine. Single-stepping the loading phase completes in a couple of minutes. Roughly the same amount of time was required to segment the traces.

From the loading phase information, we perfectly recover 46\%, 49\%, and 50\% of all the instructions in the C code, the Rust chess engine, and the Rust hash code, respectively. At least 65\% of the instructions belong to a \candidate set of size $\leq 3$ in these three programs. Only looking at the interpreter phase, we recover around 28\% of instructions with perfect information. Note that these percentages are obtained from a single execution trace and without taking into consideration the execution time of the instructions. %

\paragraph{Known Programs Classification} An application of the recovered WASM instruction traces is using it to classify which program or library is executing in the TEE among a fixed known set. For instance, this allows checking if a vulnerable version of a library is present in the confidential code supplied to the enclave. This is a useful building block for other attacks or could be used to check license violations.

Note that \ircomp is particularly vulnerable to this classification task compared to \nativecomp. This is because, in \nativecomp, they can only measure the executed instructions. This implies that the attacker would need to either know the input of the enclave or have a trace for every possible code path of the target function/library, which is being checked for presence in the enclave.
On \ircomp on the other hand the loading phase is particularly well suited to match known segments of code. This is because, generally, instructions are parsed sequentially and in the same order across executions, no matter what other inputs are provided to the enclave. Not only this but functions are also parsed independently in the WAMR loader, allowing the attacker to even check for individual matching function signatures of a library. 

We note that smaller functions are generally harder to classify than larger ones (where it is sufficient to just match with 100\% confidence a couple of marker instructions in them). We thus tested several small functions by trying to match their presence in a larger library. We took the Go Ethereum implementation\footnote{\url{https://github.com/ethereum/go-ethereum}} and compiled it to WASM. We copied the implementation of $10$ individual arithmetic functions (responsible for handling big number operations) of this project and used them in smaller programs. These smaller programs essentially simply contain a \texttt{main} function that calls the copied library functions. We then collected a trace of the loading of these small programs and segmented the WASM instructions from these traces. Finally, we tried to match the traces into a trace of the loading of the whole library. We were able to perfectly match the smaller functions in the trace of the full Go Ethereum program, thus demonstrating that segment classification is practical in the WASM system and can help us classify which WASM program is running in the enclave.

\section{Related work}\label{sec:relwork}

In this section, we discuss which side-channel attacks on TEEs we build upon and how they influence the information we assume the attacker gets access to (c.f., \cref{sec:methodology}). Note that generally these side-channels are developed to leak data from enclaves given the knowledge of the code. However, in our setting, we need to adapt them to work without any prior knowledge about the code.

\paragraph{Stack and Memory Access}
Page table-based attacks on Intel SGX exploit the untrusted OS role in managing the page tables for enclaves~\cite{sgxexplained}. The page faulting mechanism can be abused~\cite{xu2015controlled} to notify the attacker through page faults of enclave code and data accesses. Similarly, the access and dirty bits of the page table entries can be used to monitor read and writes~\cite{accesscontrolledchannel,leakycaulderon17} accesses performed by the enclave. Monitoring these bits while single-stepping gives the attacker a per-instruction resolution of these values. Moreover, the attacker can also detect control-flow changes if the instruction jumps/branches to another page. Note that these attacks are completely deterministic and noise-free.

\paragraph{Microarchitectural Structures}
Additional information can be extracted from the numerous microarchitectural details made available to the OS. While performance counters are not updated in enclave mode, their values, as measured from an attacker-controlled program, can still be influenced by the enclave execution. It is also worth mentioning the last branch record, which given knowledge of the location and target of jumps in an enclave can be used to test for branching conditions~\cite{lee2017inferring}. It is feasible to extract the LBR given the knowledge of the code, but it is challenging to employ this side-channel in our setting given that we do not know a priori the address of the jumping instructions in the confidential code.

\paragraph{Instruction Timing}
Instruction timing is considerably noisier than any previously described attack. To estimate the best resolution available to the attacker, we describe how related work leaks data from enclaves despite the noisy measurements. %
Nemesis~\cite{nemesis} observed that while single-stepping via interrupts, the interrupt delivery time is dependent on the instruction executed by the enclave. %
The attacks that leverage these timing measurements~\cite{frontal,nemesis,skarlatos2019microscope} usually perform multiple thousands of measurements for a single instruction to reduce the noise. We note that repeating measurements is not trivial and either requires the attacker's capability of re-running the enclave arbitrarily~\cite{nemesis} or a specific instruction beforehand to launch a microarchitectural replay attack~\cite{skarlatos2019microscope}. Even with the ability to repeat measurements, these attacks usually have a resolution of $40-100$ cycles.

\paragraph{Port contention}
The final source of information we consider is related to monitoring CPU port contention. Several attacks have demonstrated that port contention is a practical side-channel attack~\cite{aldaya2019port,gras2020absynthe}. However, they usually require repeated experiments to extract a signal from their noisy measurements. Nevertheless, we assume complete knowledge of the exact functional units used in the ideal attacker in \cref{sec:leakage_analysis}.

\paragraph{Summary} We chose to give the SotA attacker an even better timing resolution than what is currently feasible by allowing them a $10$ cycles resolution from a single run. Note that we also study an ideal attacker that, among other things, is cycle-accurate and can perfectly monitor the CPU port utilization. As discussed in \cref{sec:eval}, despite these capabilities, both attacker models leak very little information from the native system. On the other hand, using only controlled-channel information is enough in the WASM system to leak the vast majority of the ISA, highlighting the magnitude of the leakage amplification between the two systems.

\section{Discussion}\label{sec:discussion}

Our study considered an attacker with the goal of recovering the ISA instructions of the confidential algorithm, i.e., the opcodes.
These results can be used in different ways: we now discuss some possible practical attacks that leverage such data. %

\paragraph{Reverse-Engineering Algorithms}
A reverse engineer that wants to understand what the confidential algorithm does can leverage our results on semantically equivalent instructions (see \cref{sec:methodology}) to further reduce the number of candidate instructions and reconstruct the logic of the algorithm. Note that our attacker only leaks the instructions, but not their \textit{operands}. However, in a language like WASM, this is irrelevant for most instructions since their operands are implicit. For instance, an addition in WASM implicitly operates on the last two values present on the stack. Thus leaking that an addition was performed is enough to also leak the operands in this case. Note, however, that even in WASM, some instructions take constant values as parameters. These instructions can move values around on the stack based on their operand. We leave the task of leaking the operands for these instructions as future work.

\subsection{Applicability to Other Languages}
For our evaluation, we chose WebAssembly (WASM) as the language to instantiate the \ircomp that we studied (cf. \cref{sec:overview}).
However, some code confidentiality designs in TEE (e.g., Scone~\cite{scone}) also support different interpreted languages, e.g., Python and NodeJS.
The methods we introduced in this paper can easily be applied to analyze how much the translators of these other languages amplify the instruction leakage. As far as we are aware, their translators are not designed to provide code confidentiality, so we expect them to exhibit similar levels of leakage.

\subsection{Applicability to SEV}
To our understanding, the side-channel information and capabilities of the attacker that we use for SGX apply to SEV as well. Particularly, most of the side-channel information we use relies on manipulating interrupts and on monitoring page-level accesses. While no framework exists for SEV to conveniently replicate these functionalities, we remark that the hypervisor already performs these tasks during normal VM management. For instance, the hypervisor can schedule preemption interrupts and can tamper with page-level accesses by modifying the 2-level page translation structures. We thus conclude that the results we obtained for WASM in SGX should apply to SEV as well.

\section{Conclusions}
In this paper, we studied two different approaches commonly used for deploying confidential code into TEEs - deploying native binaries and intermediate representation (IR) - against state-of-the-art side-channel attacks. We develop a novel methodology to analyze the side-channels leakage of these approaches. We experimentally validate our methodology on nine modern microarchitectures and show that IR-based confidential code deployment amplifies any leakage found in native execution deployments. We showed that native execution results in limited leakage even against an ideal attacker, while next to no code confidentiality against a state-of-the-art attacker can be achieved when using WASM as an IR.

\defbibnote{hyperlinkinfo}{} %
\renewcommand*{\bibfont}{\small} %
\printbibliography[prenote=hyperlinkinfo]

\appendices
\crefalias{section}{appendix}
\section{x86 ISA Instruction Count}\label{sec:dataset_desc_sgx}

We focus only on the 64-bit version of the x86 architecture when creating \candidate sets. In building the \candidate sets for the microarchitectures supporting SGX and SEV, we need to account for the fact that some instructions are handled differently in these environments. Particularly, in SGX, some of the instructions are illegal and thus will never be called on bug-free enclaves. On SEV, while all instructions are allowed to execute, some will cause a hypervisor intercept, thus leaking to the attacker which instruction was executed. In the case of SGX, we never include illegal instructions in a \candidate set, while in the case of SEV, we place the intercepted instructions in \candidate sets of size 1. Next, we detail what instructions exactly end up in this special classification for the two TEEs.

\paragraph{SGX} We used the information from the Intel SDM Manual~\cite{intelSDM} Volume 3D Table 35-1 to find a criterion for instructions not allowed in SGX. To summarize, instructions with privilege level lower than 3 and instructions that perform I/O operations or that could access the segment registers are considered illegal. Note that an instruction could have an illegal version and a legal version. For instance, the \texttt{mov} instruction can write to the segment registers, and that version of the instruction is illegal.  

\paragraph{SEV} Instructions that cause a hypervisor intercept on SEV are reported in ``Table 15-7. Instruction Intercepts'' of the AMD64 Architecture Programmer’s Manual~\cite{AMD_man}. Note that there might be other conditions that cause intercepts, which might leak information to the attacker, but we only consider the instructions on that table in our calculation. Finally, the dataset we used for the Zen microarchitecture was actually obtained from information collected from a Zen+ CPU from~\cite{abel19a}. The Zen+ and Zen microarchitectures support the same x86 instructions, with the only difference being that Zen+ does not provide support for SEV (and its related instructions).

\section{SotA with different attacker cycle thresholds}\label{sec:sota_thresholds}
\begin{figure}[t!]
    \centering
    \includegraphics[width=1\linewidth]{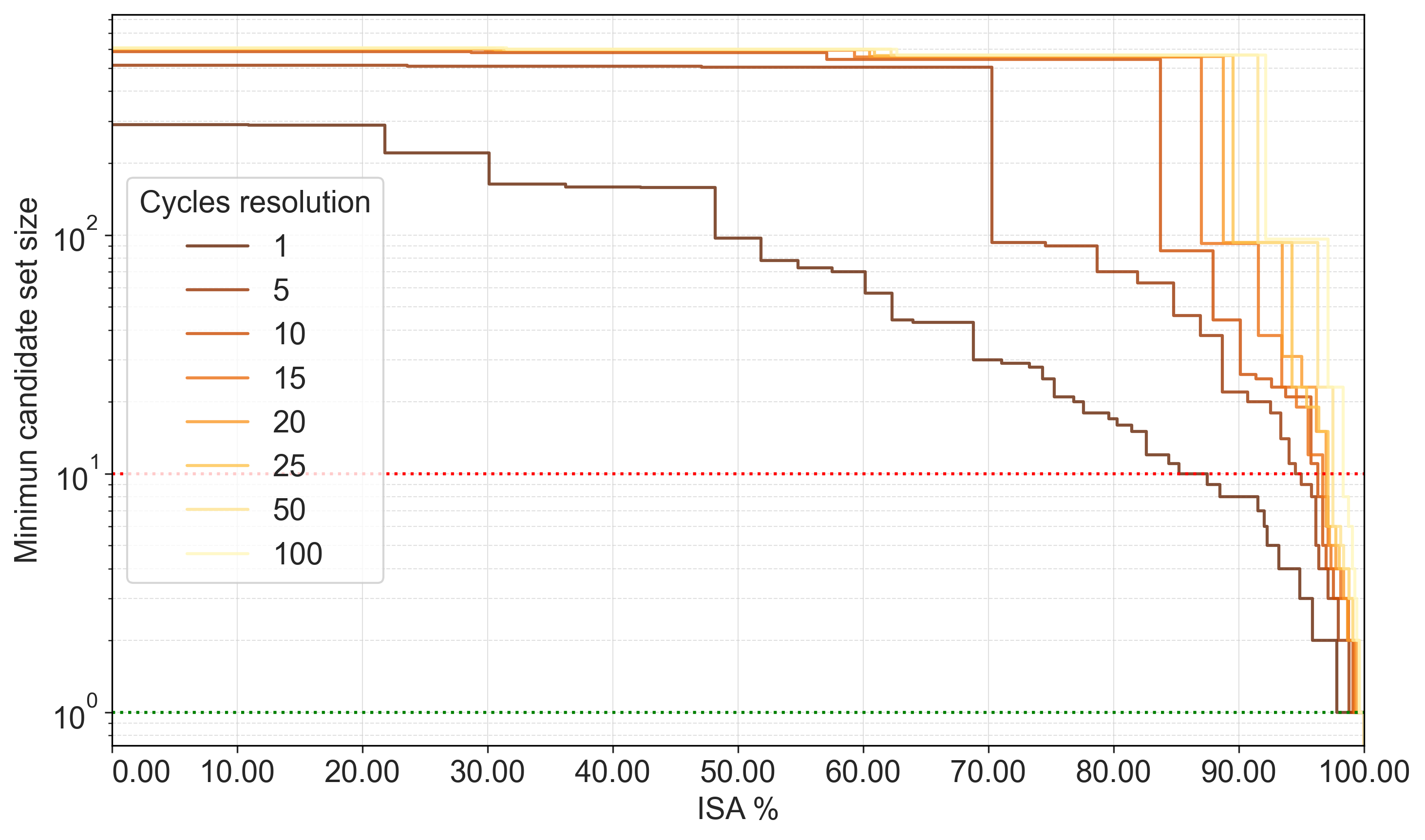}
    \vspace{-15pt}
    \caption{\Candidate set sizes distributions on the Skylake microarchitecture for an attacker in the native system with varying cycle accuracy thresholds.}
    \label{fig:cycle_relationship}
\end{figure}

To give an idea of the relationship between the strength of the attacker's instruction cycle resolution and the native system information leakage, we show in \cref{fig:cycle_relationship} how the \candidate set sizes change with different thresholds for the attacker resolution.

\end{document}